# Algebra of Deformed Differential Operators and Induced Integrable Toda Field Theory.

**I. BENKADDOUR, M. HSSAINI, M. KESSABI, B. MAROUFI**
UFR-HEP, Section Physique des Hautes Energies, Département de Physique
Faculté des sciences, B.P.1014, Rabat, Morocco
and
**M.B.SEDRA[1]**
Laboratoire de Physique Théorique et Appliquée (L.P.T.A.) Département de
Physique, Faculté des sciences, B. P.133, Kénitra, Morocco

**Abstract:**

We build in this paper the algebra of q-deformed pseudo-differential operators shown to be an essential step towards setting a q-deformed integrability program. In fact, using the results of this q-deformed algebra, we derive the q-analogues of the generalized KdV hierarchy. We focus in particular on the first leading orders of this q-deformed hierarchy namely the q-KdV and q-Boussinesq integrable systems. We present also the q-generalization of the conformal transformations of the currents $u_n, n \geq 2$ and discuss the primarity condition of the fields $w_n, n \geq 2$ by using the Volterra gauge group transformations for the q-covariant Lax operators. An induced su(n)-Toda(su(2)-Liouville) field theory construction is discussed and other important features are presented.

[1] Corresponding author: sedra@ictp.trieste.it

# 1. Introduction

An interesting subject which have been studied recently from different point of views deals with the field of non-linear integrable systems and their various higher and lower spin extensions [1,2,3,4]. These are exactly solvable models exhibiting a very rich structure in lower dimensions and are involved in many areas of mathematical physics. One recalls for instance the two dimensional Toda (Liouville) fields theories [5,2] and the KdV and KP hierarchy models [1,2], both in the bosonic as well as in the supersymmetric case.

Non linear integrable models are associated to systems of non-linear differential equations, which we can solve exactly. Mathematically these models have become more fascinating by introducing some new concepts such as the infinite dimensional Lie (super) algebras.[7], Kac-Moody algebras [8], W-algebras [3,4], quantum groups [9] and the theory of formal pseudo-differential operators [1,2,6]. Note by the way that techniques developed for the analysis of non-linear integrable systems and quantum groups can be used to understand many features appearing in various problems of theoretical physics [10,11].

Recall that, since symmetries play an important role in physics; the principal task of quantum groups consist in extending these standard symmetries to the deformed ones, which might be used in physics as well.

Motivated by the relevance of both the generalized integrable KdV hierarchies and quantum deformations, we focus in this work to present a systematic study of bidimensional q-deformed non-linear integrable models. We start then in *section 2* by presenting the algebra of q-deformed pseudo-differential operators. This provides the basic ingredients, which we need in the q-deformed integrability framework. Using these backgrounds, we will build, in *section 3*, the q-analogues of the generalized KdV hierarchy. We will concentrate in particular on the first leading orders of this hierarchy namely the q-KdV and q-Boussinesq integrable systems. In *section 4*, we present the q-generalization of the conformal transformations of the currents $u_n, n \geq 2$ and discuss the primarity condition of the fields $w_n, n \geq 2$ by using the Volterra gauge group transformations for the q-covariant Lax operators. An induced su(n)-Toda(su(2)-Liouville) field theory construction is presented in *section 5*. Other important results and some useful formulas are reported in the *appendices [A-E]*. We complete this work by a conclusion.

## 2. The Algebra of q-Deformed Pseudo-Differential Operators

We start in this section from the well known q-deformed derivation law, $\partial z = 1 + qz \partial$ [9] and derive the q-analogue of the Leibnitz rule for both local and non-local differential operators. This result, which gives naturally the algebra of q-deformed (pseudo)-differential operators, will provide a way for generating a hierarchy of q-deformed Lax evolution equations.

### *2.1 The ring of q-"analytic" currents*

To start let us precise that the deformation parameter q we consider in this study is assumed to be, a non-vanishing positive number[2]. Consider then the following q-deformed derivation rule [9]

$$\partial z = 1 + qz\partial \qquad (2.1)$$

where the symbol $\partial$ stands for the q-derivative $\partial \equiv \partial_q = (\partial/\partial z)_q$.

As we already know, "conserved" currents are ingredients that we need highly in the programs of non-linear integrable models and two-dimensional conformal field theory building. As we are interested in the present study to set-up the basic tools towards extending such programs to q-analogue ones, we will try to describe first the ring of arbitrary q-"analytic" fields which we denote by R. Following the analysis developed in[6], this space describes a tensor algebra of fields of arbitrary conformal spin. This is a completely reducible infinite dimensional SO(2) Lorentz representation (module) that can be decomposed as

$$R = \underset{k \in z}{\oplus} R_k^{(0,0)} \quad , \qquad (2.2)$$

where $R_k^{(0,0)} = R_k$ are one dimensional spin k-irreducible modules generated by the q-"analytic" fields $u_k(z)$ of «conformal» spin $k \in Z$. The upper indices (0,0) carried by R and that we shall drop whenever no confusion can arise, are special values of general indices (r,s) introduced in [6] and referring to the lowest and highest degrees of some pseudo-differential operators.

Inspiring from the derivation law Eq(2.1), we introduce in this ring a q-deformed derivative $\partial \equiv \partial_q$ satisfying

$$\partial u_k(z) = u'_k(z) + \overline{q}^k u_k(z)\partial \qquad (2.3)$$

with $\bar{q} = q^{-1}$ and $u'_k = (\frac{\partial u_k}{\partial z})_q$ stands for the standard prime derivative. Note by the way, the important fact, that we have to distinguish between the prime derivative $u'_k = (\partial u_k)$ and the operator derivative $\partial u_k = (\partial u_k) + \bar{q}^k u_k \partial$ Eq(2.3). To illustrate what does it means, consider the following examples

**1.** $u_{-k}(z) = z^k, k \geq 0.$

For this choice of the field $u_{-k}(z)$, we derive the following expression

$$(u_{-k})'(z) = (\sum_{i=0}^{k-1} q^i) z^{k-1}, \tag{2.4}$$

as we can easily check by proceeding with the first leading terms $k=0,1,2,...$

Indeed, for k=0, $(u_0)'(z) = 0$ and for k=1 we have $u_{-1} \equiv z$ and by virtue of Eq(2.1) we have

$$\begin{aligned}(u_{-1})'(z) \equiv (\partial u_{-1}) &= \partial u_{-1} - \bar{q}^{-1} u_{-1} \partial \\ &= \partial z - \bar{q}^{-1} z \partial \\ &= 1\end{aligned} \tag{2.5}$$

which we can derive also from Eq(2.4), with $\bar{q}^{-1} = q$. The non trivial case is given by k=2, such that $u_{-2} \equiv z^2$, we have

$$\begin{aligned}(u_{-2})'(z) \equiv (\partial u_{-2}) &= \partial z^2 - \bar{q}^{-2} u_{-2} \partial \\ &= (1+q)z + q^2 z^2 \partial - \bar{q}^{-2} z^2 \partial \\ &= (1+q)z\end{aligned} \tag{2.6}$$

which can also easily seen from Eq(2.4). These first leading cases, show then clearly from where the prime derivative formula Eq(2.4) comes from.

The total Leibnitz derivative applied to $u_{-k}(z) = z^k, k \geq 0$ is simply derived using successive action of the deformed q-derivative $\partial \equiv \partial_q$. We find

$$\partial z^k = (\sum_{i=0}^{k-1} q^i) z^{k-1} + q^k z^k \partial, \tag{2.7}$$

---

[2] This means that q ∈ R*. However if we suppose that q∈ C, then we shall impose for q to differ from the k-th root of unity i.e $q^k \neq 1$ as we will show for example in Eqs(2.7,8). This requirement is justified by our need of consistency when we go to the standard limit q = 1.

which justify, in some sense, the consistency of Eq.(2.4) in describing the "conformal spin" content of the analytic fields $u_k(z)$. Setting k=1, one recovers in a natural way, the standard relation Eq(2.1) just by setting k=1. The second example we consider is

**2.** $u_k(z) = z^{-k}, k \geq 1$,

Corresponding relations are computed in the same way. We find

$$\partial z^{-k} = -(\sum_{i=1}^{k} \bar{q}^i) z^{-k-1} + \bar{q}^k z^{-k} \partial \quad , \quad (2.8)$$

which reduces to

$$\partial z^{-1} = -\bar{q} z^{-2} + \bar{q} z^{-1} \partial \quad (2.9)$$

upon setting k=1.

Now having introduced the ring R, of analytic q-deformed currents, and show how the q-deformed derivative acts on; we are now in position to introduce the space of q-deformed (pseudo)-differential operators.

### *2.2 The space of q-deformed Lax operators*

Let $\Xi_m^{(r,s)}$ denote the space of q-deformed local differential operators, labelled by three quantum numbers m, r and s defining respectively the conformal spin, the lowest and the highest degrees. Typical elements of this space, are given by

$$L_m = \sum_{i=r}^{s} u_{m-i}(z) \partial^i, r, s, m \in Z \quad . \quad (2.10)$$

The symbol $\partial$ stands for the q-derivative and $u_{m-i}(z)$ are analytic fields of conformal spin (m-i). The space $\Xi_m^{(r,s)}$ behaves then as a (1+s-r)-dimensional space generated by $L_m^{(r,s)} \equiv L_m$ and whose space decomposition, is given by the linear sum

$$\Xi_m^{(r,s)} = \bigoplus_{i=r}^{s} \Xi_m^{(i,i)} \quad , \quad (2.11)$$

with

$$\Xi_m^{(i,i)} = R_m \otimes \partial^i \quad (2.12)$$

A special example of the space $\Xi_m^{(r,s)}$ is given by $R_m \equiv \Xi_m^{(0,0)}$ Eq(2.2), the set of analytic fields $u_m(z)$ introduced previously and $\partial^i \equiv \partial_q^i$ is the i-th q-derivative. A natural example of Eq (2.10), is given by the q-deformed Hill operator,

$$L_2 = \partial^2 + u_2(z) \ , \tag{2.13}$$

which will play an important role in the study of the q-deformed KdV equation and the associated "conformal" q- Liouville field theory.

A result concerning the algebra $\Xi_m^{(r,s)}$ is the derivation of the q-Leibnitz rule for local q-differential operators. Focussing to derive the general formula, let us start first by examining the first leading orders. Iteration processing applied to Eq (2.3) gives the following relations

$$\begin{aligned}
\partial u_k(z) &= u'_k(z) + \bar{q}^k u_k(z)\partial \\
\partial^2 u_k(z) &= u''_k(z) + \bar{q}^k(1+\bar{q})u'_k(z)\partial + \bar{q}^{2k} u_k(z)\partial^2 \\
\partial^3 u_k(z) &= u'''_k(z) + \bar{q}^k(1+\bar{q}+\bar{q}^2)u''_k(z)\partial + \bar{q}^{2k}(1+\bar{q}+\bar{q}^2)u'_k(z)\partial^2 + \bar{q}^{3k} u_k(z)\partial^3 \\
&\ldots
\end{aligned} \tag{2.14}$$

The crucial point was the observation that[3], these higher first order derivations formulas can be summarised into the following general Leibnitz rule

$$\partial^p u_k(z) = \sum_{j=0}^{p} \bar{q}^{(p-j)k} \chi_p^j(q) u_k^{(j)}(z) \partial^{p-j}, \ p \geq 0 \ , \tag{2.15a}$$

where $\chi_p^j(q)$ are q-coefficient functions that we have introduced such that

$$\chi_p^0(q) = \chi_p^p(q) = 1 \tag{2.15b}$$

and

$$\begin{aligned}
\chi_p^j(q) &= 1 + \bar{q}^j \sum_{m_1=0}^{j-1} q^{m_1} + \bar{q}^{2j} \sum_{m_1=0}^{j-1} \sum_{m_2=0}^{j-1-m_1} q^{2m_1+m_2} \\
&+ \bar{q}^{3j} \sum_{m_1=0}^{j-1} \sum_{m_2=0}^{(j-1-m_1)} \sum_{m_3=0}^{(j-1-(m_1+m_2))} q^{3m_1+2m_2+m_3} \\
&+ \ldots\ldots \\
&+ \bar{q}^{(p-j)j} \sum_{m_1=0}^{j-1} \sum_{m_2=0}^{j-1-m_1} \ldots \sum_{m_{p-j}=0}^{j-1-\sum_{i=1}^{p-j-1} m_i} q^{\sum_{\beta=0}^{p-j-1}(p-j-1-\beta)m_{\beta+1}} .
\end{aligned} \tag{2.15.c}$$

---

[3] This observation was possible after performing several non trivial algebraic manipulations towards wrting Eqs(2.14) into a compact form.

for $1 \leq j \leq p-1$. Some remarks are in order:

1. *From the q-Leibnitz rule Eq(2.15a), one can deduce the q-analogue of the standard binomial coefficients $c_p^j$ as follows:*

$$c_p^0 \xrightarrow{q} \bar{q}^{pk} \chi_p^0(q) \equiv \bar{q}^{pk}$$
$$c_p^p \xrightarrow{q} \chi_p^p(q) = 1$$
(2.16a)

*and for $1 \leq j \leq p-1$*

$$c_p^j \xrightarrow{q} \bar{q}^{(p-j)k} \chi_p^j(q)$$
(2.16b)

2. *Setting q=1, the local Leibnitz rule Eq(2.15a), reduces naturally to the standard derivation law:*

$$\partial^p u_k(z) = \sum_{j=0}^{p} c_p^j u_k^{(j)}(z) \partial^{p-j}, p \geq 0 \quad,$$
(2.17a)

*giving rise to the following useful relations*

$$\chi_p^0(1) = c_p^0 = 1$$
$$\chi_p^p(1) = c_p^p = 1$$
(2.17b)

*and for $1 \leq j \leq p-1$*

$$c_p^j = \chi_p^j(1) = 1 + j + \frac{j(j+1)}{2}$$
$$+ \sum_{m_1=0}^{j-1} \sum_{m_2=0}^{(j-1-m_1)} \sum_{m_3=0}^{(j-1-m_1-m_2)} 1$$
$$+ \ldots\ldots + \sum_{m_1=0}^{j-1} \sum_{m_2=0}^{j-1-m_1} \ldots\ldots\ldots \sum_{m_{p-j}=0}^{j-1-\sum_{i=1}^{p-j-1}} 1$$
(2.17c)

3. *As we can easily check, Eq(2.15c) is a sum of (p-j+1) objects starting from the value 1 which corresponds to set (j=p) with zero summation. In each term of the remaining (p-j) objects, we have a product of (n) summation $\sum_{m_1=0} \sum_{m_2=0} \ldots \sum_{m_n=0}$ with $1 \leq n \leq p-j$. This structure is useful in the standard limit q=1, recovering then the explicit form Eq(2.17c) of the well known binomial coefficient*

$$c_p^j = \frac{p!}{(p-j)! j!}$$

Moreover, Eq.(2.10) which is well defined, for local differential operators with $s \geq r \geq 0$, may be extended the negative integers(non local ones) by introducing q-deformed pseudo-differential operators of the type $\partial_q^{-p}$, $p \geq 1$, whose action on the fields $u_k(z)$ of conformal spin $k \in Z$ is constrained to satisfy:

$$\partial^p \partial^{-p} u_k(z) = \partial^{-p} \partial^p u_k(z) = u_k(z) \qquad (2.18)$$

Following the same analysis developed previously, we derive the following formulas

$$\partial^{-1} u_k(z) = \sum_{i=0}^{\infty} (-)^i q^{(k(i+1)+\frac{i(i+1)}{2})} u_k^{(i)}(z) \partial^{-i-1}$$

$$\partial^{-2} u_k(z) = \sum_{i=0}^{\infty} (-)^i q^{[k(i+2)+\frac{i(i+1)}{2}]} (\sum_{j=0}^{i} q^j) u_k^{(i)}(z) \partial^{-2-i},$$

$$\partial^{-3} u_k(z) = \sum_{i=0}^{\infty} (-)^i q^{[k(i+3)+\frac{i(i+1)}{2}]} (\sum_{j_1=0}^{i} \sum_{j_2=0}^{j_1} q^{j_1+j_2}) u_k^{(i)} \partial^{-3-i},$$

$$\ldots \qquad (2.19)$$

From these first leading formulas, we extract the following non-local Leibnitz rule

$$\partial^{-p} u_k(z) = \sum_{i=0}^{\infty} (-)^i q^{[k(i+p)+\frac{i(i+1)}{2}]} [\sum_{j_1=0}^{i} \sum_{j_2=0}^{j_1} \ldots \sum_{j_{p-1}=0}^{j_{p-2}} q^{\sum_{m=1}^{p-1} j_m}] u_k^{(i)}(z) \partial^{-p-i} \quad (2.20)$$

Here also we remark that, for a fixed value of $p \geq 1$, we have a q-deformed binomial coefficient given by a product of (p-1)-summation $\sum_{m_1=0} \ldots \sum_{m_{p-1}=0}$. Setting q=1 one recovers the standard Leibnitz rule for non local differential operators namely;

$$\partial^{-p} u_k(z) = \sum_{i=0}^{p} (-)^i C_{i+p-1}^i u_k^{(i)} \partial^{-p-1} \qquad , \qquad (2.21)$$

for $p \geq 1$, with the identity relation,

$$C_{i+p-1}^i = \sum_{j_1=0}^{i} \sum_{j_2=0}^{j_1} \ldots \sum_{j_{p-1}=0}^{j_{p-2}} 1 \qquad (2.22)$$

coinciding exactly with $\chi_{i+p-1}^i(1)$ as we can easily learn from Eq.(2.17b). Other important results are reported in the appendix A.

Up to now, we have introduced the ring R of analytic functions and construct the space of arbitrary q-deformed Lax operators by deriving the generalized q-Leibnitz rules. Next task is to see how we can apply the obtained results, to study some important features of non-linear integrable systems and conformal symmetry. Special examples namely, the Liouville field theory and the KdV equation, as well as their extensions will be considered.

### 3-Generalized q-Deformed KdV hierarchy:

We propose in this section to apply the results found previously to build the q-analogues of the KdV-hierarchy systems. We will consider in particular the first leading orders of this hierarchy namely the KdV and Boussinesq integrable systems.

Let us consider the q-deformed KdV Lax operator

$$L_2 = \partial^2 + U_2 \quad , \tag{3.1}$$

which belongs to the coset space $\Xi_2^{(0,2)} / \Xi_2^{(1,1)}$, for which we have $u_0 = 1$ and $u_1 = 0$. As known from standard references in non-linear integrable models [1,2], we can set by analogy

$$\frac{\partial L_2}{\partial t_{2n+1}} = [H_{2n+1}, L_2]_q \tag{3.2}$$

which gives the n-th evolution equation of the q-deformed KdV -hierarchy with

$$H_{2n+1} = \left( L_2^{\frac{2n+1}{2}} \right)_+ \quad . \tag{3.3}$$

The index + in Eq(3.3), stands for the local part of the deformed pseudo-differential operator $L_2^{\frac{2n+1}{2}}$ defined as

$$L_2^{\frac{2n+1}{2}} = L_2^{1/2} L_2^n \quad . \tag{3.4}$$

$L_2^{1/2}$ is just the half power of the q-KdV Lax operator introduced in Eq(3.1). It describes a q-deformed pseudo-differential operator of dimension $2 \times \frac{1}{2} = 1$. The non linear q-deformed pseudo-differential operator $L_2^{\frac{2n+1}{2}}$ is just the *(2n+1)-th* power of $L_2^{1/2}$. The standard method used to construct such kinds of operators can be found in one of the references cited in [1]. To work out explicitly $H_{2n+1}$ we need first to compute $L_2^{1/2}$. Using dimensional arguments we assume that $L_2^{1/2}$ takes the following form

$$L_2^{1/2} = \partial + a(q)u_2\partial^{-1} + b(q)u_2{}'\partial^{-2} + (c(q)u_2{}'' - d(q)u_2{}^2)\partial^{-3} + \ldots \ldots \quad (3.5)$$

where the first leading coefficients a, b, c and d are required to satisfy

$$L_2 = L_2^{1/2} L_2^{1/2} \quad . \quad (3.6)$$

Using this requirement, we find explicitly

$$a(q) = \frac{1}{1+q^{-2}}$$

$$b(q) = \frac{-1}{(1+q^{-3})(1+q^{-2})}$$

$$c(q) = \frac{1}{(1+q^{-2})(1+q^{-3})(1+q^{-4})}$$

$$d(q) = \frac{q^2}{(1+q^{-2})^2(1+q^{-4})} \quad (3.7)$$

Later on, we will introduce the dot on the analytic fields $u_2$ to describe the derivation with respect to time coordinates while the prime derivative is already introduced in Eq(2.3) to denote the derivation with respect to the space variable z.

Furthermore, the bracket introduced in Eq(3.2) is nothing but the q-deformed commutator, which we define as

$$[f\partial^n, g\partial^m]_q = \bar{q}^{m\tilde{f}} f\partial^n g\partial^m - \bar{q}^{n\tilde{g}} g\partial^m f\partial^n \quad (3.8)$$

where $f$ and $g$ are two arbitrary functions of conformal spin $\tilde{f}$ and $\tilde{g}$.

Setting n=0, Eq(3.2) becomes

$$\frac{\partial L_2}{\partial t_1} = [H_1, L_2]_q \quad (3.9)$$

where $H_1 = (L_2^{1/2})_+ = \partial$. We show also that Eq(3.9) corresponds simply to the chiral wave equation,

$$\dot{u}_2 = u_2' \quad (3.10)$$

which means the equality of dimensions $[t_1] = [z]$. For n=1, one have

$$\frac{\partial L_2}{\partial t_3} = [(L_2^{3/2})_+, L_2]_q \quad (3.11)$$

where $(L_2^{3/2})_+$, explicitly given by

$$(L_2^{3/2})_+ = \partial^3 + (\bar{q}^{-2} + a(q))u_2\partial + (1+b(q))u_2', \qquad (3.12)$$

is the q-deformed Hamiltonian operator associated with the q-Virasoro algebra.

Injecting this expression into Eq(3.11) we can extract a non linear differential equation giving the evolution of the q-spin two current $u_2$, once some easy algebraic manipulations are done. Indeed, identifying the r.h.s. and l.h.s. terms of Eq(3.11), we shall impose for some terms of the r.h.s to vanish. We obtain then the following differential equation

$$\dot{u}_2 = A(q)u_2 u_2' + B(q)u_2''' \qquad , \qquad (3.13)$$

where A(q) and B(q) are two constrained, q-dependent coefficients functions, which can be determined by a require of consistency. Simple computations lead then to

$$\begin{aligned} A(q) &= \frac{1+\bar{q}+\bar{q}^4}{1+\bar{q}^2} \\ B(q) &= -\frac{1+\bar{q}+\bar{q}^2}{(\bar{q}+1)^2} \end{aligned} \qquad , \qquad (3.14)$$

This non-linear differential equation is nothing but the q-deformed KdV system.

$$\dot{u}_2 = (\frac{1+\bar{q}+\bar{q}^4}{1+\bar{q}^2})u_2 u_2' - \frac{1+\bar{q}+\bar{q}^2}{(1+\bar{q})^2}u_2''' \qquad (3.15)$$

which coincides in the classical limit with the well known KdV integrable system [1, *A. Das*]

$$\dot{u}_2 = \frac{3}{2}u_2 u_2' - \frac{3}{4}u_2''' \qquad , \qquad (3.16)$$

and associated to the Hamiltonian differential operator

$$(L_2^{3/2})_+ = \partial^3 + \tfrac{3}{2}u_2\partial + \tfrac{3}{4}u_2' \qquad (3.17)$$

The same computations hold for the q-deformed Boussinesq equation. For more details concerning the results obtained for this system, we refer to appendix B. Note finally that the deformed KdV hierarchy discussed in this paper is based on the structure of the algebra of q-pseudo-differential operators as described previously. Other q-deformation of this hierarchy are also possible, as an example we refer the reader to following paper[13]

## 4- Conformal Transformations and q-W Currents:

We start this section by presenting the conformal transformation of the spin two current $u_2(z)$ of the q-KdV hierarchy and give later the general relations for the higher spin conformal currents $u_n(z)$, $n \geq 2$. We discuss also the primarity condition of the fields $w_n$, $n \geq 2$ by using the Volterra gauge group transformations for the q- covariant Lax operators.

### *4.1 q-generalized conformal transformations*

Let

$$L_2 = \partial^2 + u_2 \tag{4.1}$$

be the Lax operator of the q-KdV hierarchy discussed in section 3. Now we want to show how the spin 2 conformal current $u_2(z)$ transform under a conformal transformation:

$$z \to \tilde{z} = f(z). \tag{4.2}$$

Under such a transformation, we assume that the q-deformed KdV Lax operator Eq(4.1) transform as [2,12] :

$$L_2(u(z)) \to \tilde{L}_2(\tilde{u}_2(z)) = \psi^{-3/2} L_2(u_2(z)) \psi^{1/2} \tag{4.3}$$

where $\psi = \frac{\partial z}{\partial \tilde{z}}$. The choice of $\psi$ - powers in Eq(4.3) is dictated by the fact that $L_2(u(z))$ maps densities of degree $(-1/2)$ to densities of degree $(+3/2)$. We have

$$\partial \to \tilde{\partial} = \psi \partial \quad, \tag{4.4}$$

which imply

$$\tilde{\partial}^2 = \psi \psi' \partial + \psi^2 \partial^2 \quad. \tag{4.5}$$

Using straightforward computations, we find

$$\psi^{\frac{3}{2}} L_2 \psi^{\frac{1}{2}} = \psi^2 \partial^2 + \frac{1}{2}(1+\overline{q})\psi' \psi \partial + \psi^2 u_2 + \frac{1}{2}\left(\psi'' \psi - \frac{1}{2}\overline{q}(\psi')^2\right) \quad, \tag{4.6}$$

from which we can easily derive the following result.

$$\tilde{L}_2(\tilde{u}(\tilde{z})) = \tilde{\partial}^2 + \frac{\overline{q}-1}{2}\psi' \tilde{\partial} + \tilde{u}_2 \quad. \tag{4.7}$$

This shows clearly, how the conformal transformation violates the standard covariantisation property in the case of q-Lax operators. However, for q=1, we recover this property naturally, since the coefficient term of $\tilde{\partial}$ in Eq(4.7) vanishes as is proportional to $\frac{\bar{q}-1}{2}$.

Using the identification Eq(4.7), we obtain the following conformal transformation for the field $u_2(z)$

$$u_2(z) = \psi^{-2}\tilde{u}_2(\tilde{z}) - \frac{1}{2}S^{(2)}_{u_2}(q,\psi). \tag{4.8}$$

where we denote by $S^{(2)}_{u_2}(q,\psi)$ the q-Shwarzian derivative associated with the current $u_2$ and defined as :

$$S^{(2)}_{u_2}(q,\psi) = \frac{\psi''}{\psi} - \frac{1}{2}\bar{q}(\frac{\psi'}{\psi})^2. \tag{4.9}$$

The upper index (2) in $S^{(2)}_{u_2}$ stands for the order of the q-KdV hierarchy

Furthermore, Eq(4.8) shows that $u_2(z)$ transform, as a field of conformal spin two, up to an anomalous term $S^{(2)}_{u_2}(q,\psi)$ exactly like the energy momentum tensor of two dimensional conformal fields theories.

The second example we consider is the q-Boussinesq hierarchy associated with the q-deformed Lax operator

$$L_3(u_2, u_3) = \partial^3 + u_2\partial + u_3 \quad . \tag{4.10}$$

Similarly; the conformal transformation Eq(4.2) imply in this case

$$L_3(u_2, u_3) \rightarrow \tilde{L}_3(\tilde{u}_2, \tilde{u}_3) = \psi^2 L_3(u_2, u_3)\psi \quad , \tag{4.11}$$

leading to the following result

$$\psi^2 L_3 \psi = \psi^3 \partial^3 + (1+\bar{q}+\bar{q}^{-2})\psi^2\psi'\partial^2 + \{(1+\bar{q}+\bar{q}^{-2})\psi^2\psi'' + u_2\psi^3\}\partial + u_3\psi^3 + u_2\psi^2\psi' + \psi^2\psi''', \tag{4.12}$$

with

$$\tilde{L}_3(\tilde{u}_2, \tilde{u}_3) = \tilde{\partial}^3 + (\bar{q}^{-2}-1)\psi'\tilde{\partial}^2 + \tilde{u}_2\tilde{\partial} + \tilde{u}_3 \quad . \tag{4.13}$$

Using again the identification Eq(4.11), we obtain the following results

$$\begin{aligned} u_2 &= \psi^{-2}\tilde{u}_2 + S^{(3)}_{u_2}(q,\psi)(a) \\ u_3 &= \psi^{-3}\tilde{u}_3 - \frac{\psi'}{\psi}\tilde{u}_2 + S^{(3)}_{u_2}(q,\psi)(b) \end{aligned} \tag{4.14}$$

where $S_{u_2}^{(3)}$ and $S_{u_3}^{(3)}$ are the q-Shwarzian derivatives associated respectively with the conformal $u_2$ and $u_3$. They are given by

$$S_{u_2}^{(3)}(q,\psi) = \bar{q}^{-2}(\frac{\psi'}{\psi})^2 - \bar{q}(\bar{q}+1)\frac{\psi''}{\psi}$$
$$S_{u_3}^{(3)}(q,\psi) = \frac{\psi'''}{\psi} + \frac{\psi'}{\psi} S_{u_2}^{(3)}(q,\psi). \tag{4.15}$$

Note by the way that $S_{u_2}^{(3)}$ and $S_{u_3}^{(3)}$ are shown to relate as follows

$$\partial S_{u_2}^{(3)} + \bar{q}(\bar{q}+1) S_{u_3}^{(3)} = 0 \tag{4.16}$$

As suspected for q=1, one find the standard formulas given by [2,12]

$$u_2 = \psi^{-2}\tilde{u}_2 - S_{u_2}^{(3)}(1,\psi)$$
$$u_3 = \psi^{-3}\tilde{u}_3 - \frac{\psi'}{\psi^3}\tilde{u}_2 - S_{u_3}^{(3)}(1,\psi), \tag{4.17}$$

with

$$S_{u_2}^{(3)}(1,\psi) = (\frac{\psi'}{\psi})^2 - 2\frac{\psi''}{\psi}$$
$$S_{u_3}^{(3)}(1,\psi) = \frac{\psi'''}{\psi} + \frac{\psi'}{\psi} S_{u_2}^{(3)}(1,\psi), \tag{4.18}$$

and

$$\partial S_{u_2}^{(3)} + 2 S_{u_3}^{(3)} = 0 \tag{4.19}$$

The presence of the anomalous term in Eq(4.14.b) can be removed away by a convenient basis choice namely the primary basis which we will discuss later on.

Having given explicitly the conformal transformation of the currents $u_2$ and $u_3$ of conformal spin 2 and 3, now we focus to generalize these results to higher conformal spin currents $u_n(z)$, n=2, 3...

Let

$$L_n[u] = \partial^n + \sum_{i=0}^{n-2} u_{n-i} \partial^i, \tag{4.20}$$

be the higher order Lax operator involving (n-1) conformal currents with $u_0 = 1$ and $u_1 = 0$ and where $\partial = \partial_q$. Under the conformal transformation Eq(4.2); this Lax operator is assumed to transform as :

$$L_n[u] \to \widetilde{L}_n[\widetilde{u}] = \psi^{\frac{n+1}{2}} L_n[u] \psi^{\frac{n-1}{2}} \quad . \tag{4.21}$$

Similarly to the previous study; the structure of the Lax operator $L_n[u]$ Eq(4.20) is broken under the conformal transformation, we find in general

$$\widetilde{L}_n[\widetilde{u}] = \widetilde{\partial}^n + A\psi \widetilde{\partial}^{n-1} + \sum_{i=0}^{n-2} \widetilde{u}_{n-i} \widetilde{\partial}^i \tag{4.22}$$

where A is an arbitrary Lorentz scalar function which we will precise.

To determine $\widetilde{L}_n$, we need to compute explicitly $\widetilde{\partial}^n$. Starting from Eq(4.4) and using simply algebraic manipulation, we find the following results:

$$\widetilde{\partial}^n = \sum_{i=1}^{n} M_i^n \partial^i \quad , \tag{4.23}$$

where $M_i^n$; are functions of conformal spin (n-i), which we can summarise as follows

$$\begin{aligned} M_n^n &= \psi^n \\ M_1^n &= \psi \partial \, M_1^{n-1} \\ M_i^n &= \psi \left[ M_{i-1}^{n-1} \overline{q}^{(n-i)} + \partial M_i^{n-1} \right] \quad 2 \leq i \leq n-1 \end{aligned} \tag{4.24}$$

Substituting these relations into Eq(4.22) we find

$$\widetilde{L}_n = \sum_{i=0}^{n} X_i(A, M, \psi) \partial^{n-i} \quad , \tag{4.25}$$

where

$$X_i(A, M, \psi) = \sum_{j=0}^{i} \widetilde{u}_j M_{n-i}^{n-j} \quad . \tag{4.26}$$

In the other hand; simply algebraic calculations show that the r.h.s. of Eq(4.21) lead to

$$\psi^{\frac{n+1}{2}} L_n[u] \psi^{\frac{n-1}{2}} = \psi^{\frac{n+1}{2}} \sum_{i=0}^{n} \left( \sum_{j=0}^{i} u_{i-j} X_{n-j+i}^{j}(q) \ (\psi^{\frac{n-1}{2}}) \right) \partial^{n-i} \quad , \tag{4.27}$$

Identifying then Eq(4.25) with Eq(4.27) we find:

$$A(q,\psi) = \frac{\psi^{1-n}}{\psi'}\left[\chi_n^1(q)\psi^{\frac{n+1}{2}}(\psi^{\frac{n-1}{2}})' - M_{n-1}^n\right], \qquad (4.28)$$

with

$$\chi_n^1(q) = \sum_{i=0}^{n-1} \bar{q}^i \qquad (4.29)$$
$$\chi_n^0 = \chi_n^n = 1,$$

and

$$u_i = \psi^{-n}\left\{M_{n-i}^n + \sum_{j=1}^{i}\left[\tilde{u}_j M_{n-i}^{n-j} - \psi^{\frac{n+1}{2}} u_{i-j}\chi_{n-i+j}^j(q)\ (\psi^{\frac{n-1}{2}})^{(j)}\right]\right\} \text{ for } 0 \le i \le n. \qquad (4.30)$$

We show then clearly how transform the conformal currents $u_i, i \ge 2$ under Eq(4.2). The first thing we learn from these results is the dependence on the q-parameter, which once it coincides with q=1 lead to the standard formulas. To illustrate the obtained results; we consider two particular examples discussed previously, namely the q-KdV and q-Boussinesq integrable models described respectively by $L_2(u)$ and $L_3(u)$.

The former is easily obtained by setting n=2 into the Eqs(4.28-30) which recover Eqs(4.8-9) exactly with

$$A = \frac{\bar{q}-1}{2}. \qquad (4.31)$$

Similarly; Eqs(4.14-15) are obtained by setting n=3 Eqs(4.27-28) with

$$A = \bar{q}^2 - 1. \qquad (4.32)$$

### *4.2 Volterra gauge group transformation and q-W-currents:*

In the framework to generalize the conformal transformations to the q-deformed case, we found, in addition to new features, the presence of anomalous terms at the level of the conformal current $u_3, u_4 ... u_n$ Eq(3.30).

Our idea is to consider a Volterra gauge group transformation associated to an «orbit» in which no such anomalous terms can appear. We start first by recalling the Volterra gauge group symmetry. This is a symmetry group whose typical elements are given by the Lorentz scalar q-pseudo-differential operators [14]

$$K[a] = 1 + \sum_{i \ge 1} a_i(z)\partial^{-i}, \qquad (4.33)$$

where $a_i(z)$ are arbitrary analytic functions of conformal spin $i = 1,2,3,...$ These functions; to which we shall refer hereafter to as the Volterra gauge parameters, can be expressed in term of the residue operation as

$$a_i(z) = \text{Res}\left(K(a)\partial^{i-1}\right), \qquad (4.34a)$$

where

$$\text{Res}\,\partial^i = \delta_{i+1,0}, \qquad (4.34b)$$

and for a given function f(z), we recall that we have Eq(2.3)

$$\partial f(z) = f'(z) + \overline{q}^{\widetilde{f}} f(z) \partial \qquad (4.35)$$

Next, we will apply this Volterra gauge group symmetry to the algebra of q-Lax operators Eq(4.20) via the following relation

$$L_n(u) \to L_n(w) = K^{-1}(a) L_n(u) K(a) \qquad (4.36)$$

where $L_n(w)$ is the transform of $L_n(u)$ under the Volterra group action with $w = w(a,u)$ is a function which depends on the Volterra parameter $a_i$ and the u-fields. Moreover, Eq(4.36) shows that the u-currents may be expressed completely in terms of the Volterra gauge parameters $a_i$ and their k-th derivatives. However solving Eq(4.36); one find that the new fields $w_i$ are polynomials in the old u-fields and the Volterra parameters and their derivatives.

Making appropriate choices of the Volterra parameters dictated by the primarity condition, the w-fields can then be expressed in terms of the u-fields exactly as do the primary w-currents which satisfy [12]

$$w_s(z) = \psi^{-s} \widetilde{w}_s(\widetilde{z}) \qquad (4.37)$$

To illustrate how things work, let us focus to solve Eq(4.36) for the special case n=3. We have

$$L_3(u) = \partial^3 + u_2 \partial + u_3, \qquad (4.38)$$

describing the Lax operator of the q-Boussinesq integrable system. Applying the Volterra gauge group symmetry Eq(4.36) to Eq(4.38), by identifying

$$K(a) L_3(w) = L_3(u) K(a), \qquad (4.39)$$

we find after straightforward algebraic calculations the following formulas for the first parameters $a_1, a_2, a_3, a_4$

$$\bar{q}^3 a_1 = a_1$$
$$a_2 + w_2 = u_2 + \bar{q}^6 a_2 + \bar{q}^2(1+\bar{q}+\bar{q}^2)a_1'$$
$$a_3 + w_3 + q^2 a_1 w_2 = u_3 + \bar{q}^9 a_3 + \bar{q} a_1 u_2 + \bar{q}^4(1+\bar{q}+\bar{q}^2)a_2' + \bar{q}(1+\bar{q}+\bar{q}^2)a_1'' \quad (4.40)$$
$$a_4 + q^3 a_1 w_3 - q^5 a_1 w_2' + q^4 a_2 w_3 = a_1 u_3 + \bar{q}^2 a_2 u_2 + \bar{q}^{12} a_4 + \bar{q}^2(1+\bar{q}+\bar{q}^2)a_2' +$$
$$\bar{q}^6(1+\bar{q}+\bar{q}^2)a_3' + a_1''' + a_1' u_2.$$

We show also that the remaining Volterra parameters $a_j$, $j \geq 2$ are constrained to satisfy:

$$a_{j+3}(\bar{q}^{3(j+3)} - 1) = a_1(-1)^{j-1} q^{3j+i(\frac{j-1}{2})} w_3^{(j-1)} + a_1(-1)^j q^{2(j+1)+j(\frac{j+1}{2})} w_2^{(j)}$$
$$+ \sum_{i=0}^{\infty} a_{j-i} q^{3j+i(\frac{i+1}{2})} \left( \sum_{k_1=0}^{i} \cdots \sum_{k_{j-i-1}=0}^{k_{j-i-2}} q^{\sum_{m=1}^{j-i-1} k_m} \right) w_3^{(i)}$$
$$+ \sum_{i=0}^{\infty} a_{j-i+1} q^{2(j+1)+i(\frac{i+1}{2})} \left( \sum_{k_1=0}^{i} \cdots \sum_{k_{j-i}=0}^{k_{j-i-1}} q^{\sum_{m=1}^{j-1} k_m} \right) w_2^{(i)} \quad (4.41)$$
$$- \bar{q}^{j+1}(1+\bar{q}+\bar{q}^2)a_{j+1}'' - \bar{q}^{j+1} a_{j+1} u_2 - a_j'''$$
$$- \bar{q}^{2(j+2)}(1+\bar{q}+\bar{q}^2)a_{j+2}' - a_j' u_2 - a_j u_3,$$

Consequently, we learn from Eq(4.40) that the spin one Volterra gauge parameter $a_1$ vanishes naturally for an arbitrary values of the parameter q. This lead to set

$$\begin{aligned} a_1 &= 0 & (a) \\ (1-\bar{q}^6)a_2 &= u_2 - w_2 & (b) \\ (1-\bar{q}^9)a_3 &= u_3 - w_3 + \bar{q}^4(1+\bar{q}+\bar{q}^2)a_2' & (c) \\ (\bar{q}^9 - 1)a_4 &= q^4 a_2 w_3 - \bar{q}^2(1+\bar{q}+\bar{q}^2)a_2'' \\ &\quad - \bar{q}^6(1+\bar{q}+\bar{q}^2)a_3' - (a_1' + \bar{q}^2 a_2)u_2 & (d) \end{aligned} \quad (4.42)$$

with the constraints Eqs(4.41). Note by the way that when q=1, one recover, from Eqs(4.42), a Volterra gauge orbit $K_{q=1}\{a_i\}$ in which the $w_i$-fields are seen as primary currents [14].

Actually, our principal task is to make an appropriate choice on the Volterra parameters $a_i$ such that $w_i$ become primary conformal currents satisfying Eq(4.37). Recall also that, in the classical limit, the analytic field $u_2$ behaves as a spin 2-field of 2d conformal field theory coinciding with the $w_2$ current. Similarly, in the deformed case; we can require for $w_2$ to be proportional to $u_2$, which leads from Eq(4.42.b) to set

$$a_2 = \delta\, u_2 \quad , \tag{4.43}$$

where $\delta$ is an arbitrary constant for the moment. We have then

$$w_2 = u_2(1 - \delta(1 - \bar{q}^{-6})) \quad . \tag{4.44}$$

Plaguing Eq(4.43) into Eq(4.42.c); we obtain

$$a_3 = \beta_1 u_3 + \beta_2 u_2' \quad , \tag{4.45}$$

where $\beta_1$ and $\beta_2$ are for instance arbitrary constants which can be fixed.

The resulting expression for the q-deformed w-current of spin 3 is

$$w_3 = u_3\left[1 + (\bar{q}^9 - 1)\beta_1\right] + u_2'\left[\bar{q}^{-4}(1 + \bar{q} + \bar{q}^2)\delta + \beta_2(\bar{q}^{-9} - 1)\right] \quad , \tag{4.46}$$

with the constraints equation Eq(4.41) giving the remaining Volterra parameters $a_j$, $j \geq 5$.

$$a_4(\bar{q}^9 - 1) = q^4 a_2 w_3 - \bar{q}^2(1 + \bar{q} + \bar{q}^2)a_2'' - \bar{q}^6(1 + \bar{q} + \bar{q}^2)a_3' - (a_1' + \bar{q}^2 a_2)u_2$$

$$a_{j+3}(\bar{q}^{3(j+3)} - 1) = \sum_{i=0}^{\infty} a_{j-i} q^{3j+i(\frac{i+1}{2})} \left( \sum_{k_1=0}^{i} \cdots \sum_{k_{j-i-1}=0}^{k_{j-i-2}} q^{\sum_{m=1}^{j-i-1} k_m} \right) w_3^{(i)}$$

$$+ \sum_{i=0}^{\infty} a_{j-i+1} q^{2(j+1)+i(\frac{i+1}{2})} \left( \sum_{k_1=0}^{i} \cdots \sum_{k_{j-i}=0}^{k_{j-i-1}} q^{\sum_{m=1}^{j-i} k_m} \right) w_2^{(i)} \tag{4.47}$$

$$- \bar{q}^{j+1}(1 + \bar{q} + \bar{q}^2)a_{j+1}'' - \bar{q}^{j+1} a_{j+1} u_2 - a_j'''$$

$$- \bar{q}^{2(j+2)}(1 + \bar{q} + \bar{q}^2)a_{j+2}' - a_j' u_2 - a_j u_3$$

Now, let us consider a conformal transformation of the spin -3 w-current Eq(4.46), we find:

$$\widetilde{w}_3 = \psi^3 w_3 + y_3 \tag{4.48}$$

where $y_3$ is a function of conformal spin 3 given by

$$y_3 = \psi^2 \psi' \{1 + 2\bar{q}^{-4}(1 + \bar{q} + \bar{q}^2)\delta + (\bar{q}^9 - 1)\beta_1 + 2(\bar{q}^{-9} - 1)\beta_2\}u_2 +$$

$$+ \psi^3 \{ \left(s_{u_3}^{(3)} - \frac{\psi'}{\psi} s_{u_2}^{(3)}\right) - \left( 2\frac{\psi'}{\psi} s_{u_2}^{(3)} + \partial s_{u_2}^{(3)} \right)\bar{q}^{-4}(1 + \bar{q} + \bar{q}^2)\delta +$$

$$+ (\bar{q}^9 - 1)\left(s_{u_3}^{(3)} - \frac{\psi'}{\psi} s_{u_2}^{(3)}\right)\beta_1 - (\bar{q}^{-9} - 1)\left( 2\frac{\psi'}{\psi} s_{u_2}^{(3)} + \partial s_{u_2}^{(3)} \right)\beta_2 \}$$

Imposing the primarity condition Eq(4.37) imply the vanishing of $y_3$ from which one can derive a solution for the constants $\delta(q)$, $\beta_1(q)$ and $\beta_2(q)$ which are required to coincide in the classical limit with $\delta(1) = -1/6$, $\beta_1(1) = 0$ and $\beta_2(1) = 1/6$ respectively.

## 5- Note on the SU(n)-Toda field theory construction

The aim of this section is to set up some crucial ingredients towards building the q-deformed analogue of 2d su(n)-Toda like conformal field theory, using the previous analysis. The starting point consist in exploiting the correspondence which exist between the second Hamiltonian structure of integrable systems and the Virasoro conformal algebra which is the symmetry of 2d Liouville field theory.

Consider then, the integrable q-KdV equation, discussed previously in *Section 3* and which we can conveniently take as follows *(see Eq(3.15))*:

$$\dot{u}_2 = \left(\frac{1+\bar{q}+\bar{q}^4}{1+\bar{q}^2}\right) u_2 u_2' - \frac{1+\bar{q}+\bar{q}^2}{(\bar{q}+1)^2} u_2''' \quad . \tag{5.1}$$

Applying the Miura transformation (which connects the dynamical current $u_2$ with the scalar field $\varphi \equiv \varphi(z,\bar{z})$) to the q-deformed KdV Lax operator as follows:

$$L_2 = (\partial^2 + u_2) = (\partial + A)(\partial + B) \quad , \tag{5.2}$$

where A and B are spin 1 fields, which are constrained to satisfy

$$\begin{cases} A = -\bar{q}B \\ AB + B' = u_2 \end{cases}, \tag{5.3}$$

with $B' = (\partial B)$. A solution to this system is

$$\begin{cases} A = -\partial\varphi \\ B = q\partial\varphi \end{cases}, \tag{5.4}$$

which gives

$$u_2 = q\left(\partial^2\varphi - (\partial\varphi)^2\right) \quad . \tag{5.5}$$

This equation shows that $u_2$ is a q-deformed spin two current, which behaves like the stress-energy momentum tensor of 2d Liouville conformal field theory. An important point is to look for the lagrangian of this theory. Using standard knowledge on conformal Liouville field theory[5, 10], we can set by analogy

$$S[\varphi] = \int d^2z \left\{ \partial\varphi\bar{\partial}\varphi + \frac{2}{b}\exp(b\varphi) \right\} \tag{5.6}$$

where the coefficient number b is shown to take the value $b = (1+\bar{q})$ *(See Appendix D)*. We show also that the equation of motion which emerge from this action is nothing but the q-deformed 2d conformal Liouville equation given by

$$\partial\bar{\partial}\varphi - 2\bar{q}e^{(1+\bar{q})\varphi} = 0 \quad , \quad (\bar{q} = q^{-1}) \tag{5.7}$$

To obtain this equation, one must precise, as explicitly shown in *Appendices C* and *D* that the Euler-Lagrange equations, should be applied taking into account the previous analysis.. The q - deformed form of the conserved current can be written as

$$T(\varphi) \equiv q\partial^2\varphi - q(\partial\varphi)^2 \quad , \tag{5.8}$$

whose conservation is assured by the equation of motion Eq(5.7)

$$\bar{\partial}T(\varphi) = 0 \tag{5.9}$$

Note by the way that this conservation law combined with Eq(5.7) which fixes the q-coefficient number $b = (1+\bar{q})$ in the exponential Eq(5.6). Before closing this discussion; some remarks are in order.

Note first that the action Eq(5.6) is conformally invariant and generalize naturally the su(2) standard Liouville theory . As already known from the standard studies; the coefficient number in the exponential Liouville potential is closely connected with the Cartan matrix of some simple Lie algebra. An important task is to look for the interpretation of the coefficient $(\bar{q}+1)$, appearing in our exponential, from the Lie algebraic point of view. Remark that this number coincide in the classical limit case with the number 2 which is nothing but the Cartan matrix of the su(2) Lie algebra .

However; the choice of our q-KdV Lax operator in Eq(5.2), shows already the existence of an su(2) symmetry; which can be recovered also from the Liouville action. Indeed; if we redefine the scalar field φ to be

$$\Phi = \frac{\bar{q}+1}{2}\varphi \quad , \tag{5.10}$$

we can easily read the su(2) symmetry from the Liouville action. The latter becomes:

$$S[\Phi] = \int d^2z \left\{ \lambda\partial\Phi\bar{\partial}\Phi + \frac{2}{\bar{q}+1}\exp(2\Phi) \right\} \tag{5.11}$$

upon introducing a parameter λ namely

$$\lambda = \left(\frac{\bar{q}+1}{2}\right)^2 \qquad (5.12)$$

The q-deformed Liouville equation of motion becomes

$$\partial\bar{\partial}\Phi - \bar{q}(\bar{q}+1)\exp(2\Phi) = 0 \qquad (5.13)$$

We can also think to generalize the above q-deformed su(2)-Liouville field theory to the $su(n)$ conformal Toda field theory. We set for the moment

$$S_{su(n)Toda} = \int \partial^2 z \left( \partial\phi\bar{\partial}\phi + \eta(q)\sum_{i=1}^{n-1}\exp(\alpha_i\phi) \right) \qquad , \qquad (5.14)$$

where $\phi = \sum_{j=1}^{n-1}\alpha_j\phi_j$ and $\alpha_j$ are the simple root of the Lie algebra su(n) whose Cartan matrix is defined as

$$K_{ij} = \alpha_i \alpha_j \qquad , \qquad (5.15)$$

and where η(q) is a function of the parameter q which can be easily fixed given the corresponding model in the generalized KdV hierarchy. More on this q-deformed Toda filed theory construction may be a subject of future works

## 6- Conclusion

We tried in this work, to understand the behaviour of 2d non-linear integrable systems in the q-deformed case. For this reason, we started by generalizing some well-known results in the theory of formal pseudo-differential operators to the q-deformed case. The obtained results, are applied to build the q-analogues of the generalized integrable q-KdV hierarchies whose first leading orders are the q-KdV and q-Boussinesq systems. We derived the dynamical equations of these deformed integrable hierarchies, leading in fact to the standard ones, once the q-parameter is fixed to be one. We discussed previously, how transform in the deformed case; the currents $u_j(z)$ under a conformal transformation. The results obtained; showed a non trivial behaviour of these currents; which coincides naturally with the standard results upon setting q=1. We discussed also the primarity condition of these currents using the Volterra gauge group symmetry. In the last part of this work, devoted to the Toda field theory construction; we presented the q-analogue of the su(2) Liouville and su(n) Toda conformal field theories. Other algebraic properties are reported to appendices.

# Appendix A

Let f(z) be an arbitrary analytic function of conformal spin $\Delta f = \tilde{f}$. Using Eq(2.3) and iterative action of the q-deformed derivative we find

$$\partial f^n(z) = (1 + \bar{q}^{-1\tilde{f}} + \bar{q}^{-2\tilde{f}} + \ldots \bar{q}^{-(n-1)\tilde{f}}) f' f^{n-1} + \bar{q}^{-n\tilde{f}} f^n \partial \quad , \tag{A.1}$$

where n is a positive integer number. Setting q=1 one recovers, once again, the ordinary derivation rule,

$$\partial f^n(z) = n f' f^{n-1} + f^n \partial \quad . \tag{A.2}$$

A special choice of f(z) in Eq(A.1) is given by f(z)=z with $\tilde{z}=-1$, we have .

$$\partial z^n = (1 + q + q^2 + \ldots + q^{n-1}) z^{n-1} + q^n z^n \partial \quad , \tag{A.3}$$

which reduces to Eq.(2.1) upon setting n=1. For negative integer numbers we find easily,

$$\partial f^{-n}(z) = -(q^{\tilde{f}} + q^{2\tilde{f}} + \ldots + q^{n\tilde{f}}) f' f^{-n-1} + q^{n\tilde{f}} f^{-n} \partial \quad , \tag{A.4}$$

which becomes upon setting q=1,

$$\partial f^{-n}(z) = -n f' f^{-n-1} + f^{-n} \partial \quad . \tag{A.5}$$

As before, setting f(z)=z we obtain

$$\partial z^{-n} = -(\bar{q} + \bar{q}^2 + \ldots \bar{q}^n) z^{-n-1} + \bar{q}^n z^{-n} \partial \quad . \tag{A.6}$$

Furthermore, we note that for half integer powers of f(z) we can obtain general formulas. The method to do this start from setting

$$\partial f^{1/2} = \alpha(q) f' f^{-1/2} + \beta(q) f^{1/2} \partial \quad , \tag{A.7}$$

where $\alpha(q)$ and $\beta(q)$ are two arbitrary q-dependent functions that we can determine explicitly by the following trivial property

$$\partial (f^{1/2} f^{1/2}) \equiv \partial(f) \quad . \tag{A.8}$$

General formulas are given by:

$$\partial f^{\frac{2n+1}{2}}(z) = \frac{(1 + \bar{q}^{-\tilde{f}/2} + \bar{q}^{-2\tilde{f}/2} + \ldots + \bar{q}^{-2n\tilde{f}/2})}{(1 + \bar{q}^{\tilde{f}/2})} f' f^{\frac{2n-1}{2}} + \bar{q}^{-(2n+1)\tilde{f}/2} f^{\frac{(2n+1)}{2}} \partial \quad , \tag{A.9}$$

and

$$\partial f^{\frac{-(2n+1)}{2}} = \frac{-q^{\tilde{f}/2} (q^{\tilde{f}/2} + q^{\tilde{f}} + q^{3\tilde{f}/2} + \ldots q^{(2n+1)\tilde{f}/2})}{(1 + q^{\tilde{f}/2})} f' f^{\frac{-(2n+3)}{2}} + q^{(2n+1)\tilde{f}/2} f^{\frac{-(2n+1)}{2}} \partial \quad . \tag{A.10}$$

These q-generalized results are important in discussing the q-deformed Lax evolution equations and the covariantisation of q-differential Lax operators.

Before closing this appendix, note that the ring $R = \bigoplus_{k \in Z} R_k$ defined in Eq(2.2) is a commutative ring, which means that for each $u_k(z)$ and $u_l(z)$ belonging to R we have $u_k(z)u_l(z) = u_l(z)u_k(z)$. However, applying the q-Leibnitz rule Eq(2.3), one can easily show the existence of a non commutative structure in the space $\Xi_m^{(r,s)}$ of local and non local q-differential operators. Indeed, let f and g be two arbitrary functions of conformal spin $\tilde{f}$ and $\tilde{g}$, with $fg = gf$, we have

$$(\partial f)g = f'g + \bar{q}^{\tilde{f}} fg' + \bar{q}^{(\tilde{f}+\tilde{g})} fg\partial \quad , \tag{A.11}$$

while

$$(\partial g)f = g'f + \bar{q}^{\tilde{g}} gf' + \bar{q}^{(\tilde{f}+\tilde{g})} gf\partial , \tag{A.12}$$

which shows clearly that $(\partial f)g \neq (\partial g)f$ for $\tilde{f} \neq \tilde{g}$. Note that this non commutativity property of $f$ and $g$, with respect to the action of the q-derivative $\partial_q$, arise naturally from Eq(2.3). Note also the important fact that when the function $g$ is, for example, the n-th power of the function $f$ with $n \in R$; one can set $g = f^n$ which yields $\tilde{g} = n\tilde{f}$ and then

$$(\partial f)g = (\partial g)f \quad , \tag{A.13}$$

with $f'f^n = f^n f'$. One can then deduce that the two subspaces $R_{\tilde{f}}$ and $R_{\tilde{g}}$, of analytic functions f(z) and g(z) of conformal spin $\tilde{f}$ and $\tilde{g}$, respectively, don't commute under the action of the q-derivative $\partial_q$ unless if there exist a relative integer $n \in Z$, such that $g = f^n$.

## Appendix B: *q-deformed Boussinesq equation*:

Using the same technique developed for the q-deformed KdV system, we present in this appendix a q-generalization of the Boussinesq integrable hierarchy, *for a review see A. Das in*[1].

Let

$$L_3 = \partial^3 + u_2\partial + u_3 \tag{B.1}$$

be the Lax operator associated with the q-Boussinesq hierarchy with $u_2$ and $u_3$ are two currents of conformal spin 2 and 3, respectively. Knowing that $(L_3^{1/3})^3 = L_3$ and the fact that $L_3^{1/3}$ is an object of conformal spin 1, we can set:

$$L_3^{1/3} = \partial + au_2\partial^{-1} + (bu_3 - cu_2')\partial^{-2} + (du_2'' - eu_2^2 - fu_3')\partial^{-3} + \ldots \quad (B.2)$$

where the coefficients $a, b, c, d$ and $e$ are given explicitly by

$$a = \frac{1}{1+\bar{q}^2+\bar{q}^4}$$

$$b = \frac{1}{1+\bar{q}^3+\bar{q}^6}$$

$$c = \frac{1+\bar{q}^2+\bar{q}^3}{(1+\bar{q}^2+\bar{q}^4)(1+\bar{q}^3+\bar{q}^6)}$$

$$d = \frac{1}{(1+\bar{q}^2+\bar{q}^4)(1+\bar{q}^4+\bar{q}^8)}\left\{\frac{(1+\bar{q}^2+\bar{q}^3)(1+\bar{q}^3+\bar{q}^4)}{1+\bar{q}^3+\bar{q}^6} - 1\right\} \quad (B.3)$$

$$e = \frac{1+q^2+\bar{q}^2}{(1+\bar{q}^2+\bar{q}^4)^2}$$

$$f = \frac{1+\bar{q}^3+\bar{q}^4}{1+\bar{q}^3+\bar{q}^6}$$

so that

$$(L_3^{1/3})_+ = \partial . \quad (B.4)$$

Identifying the r.h.s. and l.h.s. of the following equation

$$\frac{\partial L_3}{\partial t_1} = [(L_3^{1/3})_+, L_3]_q \quad , \quad (B.5)$$

we obtain

$$u'_2 = \dot{u}_2$$
$$u'_3 = \dot{u}_3 \quad , \quad (B.6)$$

which give the chiral wave equations for the Boussinesq hierarchy.

Similarly, the identification:

$$\frac{\partial L_3}{\partial t_2} = [(L_3^{2/3})_+, L_3]_q \quad (B.7)$$

with:

$$(L_3^{2/3})_+ = \partial^2 + a(\bar{q}^2+1)u_2 \quad , \quad (B.8)$$

gives:

$$\dot{u}_3 = u_3'' + a(1+\bar{q}^{-2})\{\alpha u_2''' + \beta u_2 u_2'\} \quad (a)$$
$$\dot{u} = u_2''\{1+\alpha a \bar{q}^{-2}(1+\bar{q}^{-2})(1+\bar{q}+\bar{q}^{-2})\} + \bar{q}^{-3}(1+\bar{q})u_3' \quad (b) \quad (B.9)$$
$$\bar{q}^{-3}u_3'' = (1+\bar{q})\{1+\alpha a \bar{q}^{-2}(1+\bar{q}^{-2})\}u_2''' \quad (c)$$

where $\alpha$ and $\beta$ are two arbitrary functions of the parameter q which can be conveniently fixed in such way that $\alpha = \beta = -1$ in the classical limit.

Combining (B9.a) and (B9.c) we find :

$$\dot{u}_3 = -q^2(1+q+a\alpha\bar{q}(1+\bar{q}^2))u_2''' + a\beta(1+\bar{q}^2)u_2 u_2' \quad . \quad (B.10)$$

Moreover, note that (B9.c) can be written as :

$$u_3'' = \frac{-\bar{q}^3}{(1+\bar{q})(1+a\alpha\bar{q}^2(1+\bar{q}^2))} \quad , \quad (B.11)$$

which imply by virtue of (B9.b)

$$\dot{u}_2 = B(q,\alpha)u_3' \quad , \quad (B.12)$$

with :

$$B(q,\alpha) = \bar{q}^3(1+\bar{q}) - \frac{\bar{q}^3(1+a\alpha\bar{q}^2(1+\bar{q}^2)(1+\bar{q}+\bar{q}^2))}{(1+\bar{q})(1+a\alpha\bar{q}^2(1+\bar{q}^2))} \quad (B.13)$$

Equation (B10) and (B12) give then the q- deformed Boussinesq equations. Setting q=1 we recover the classical Boussineq equation namely [*A. Das,* 1 ]

$$\dot{u}_2 = \tfrac{7}{2} u_3'$$
$$\dot{u}_3 = -\tfrac{4}{3} u_2''' - \tfrac{2}{3} u_2 u_2'. \quad (B.14)$$

Next we will show how the q- deformed Boussinesq equation (B10, 12) can be cast into a simple form. Indeed using straightforward algebraic computations (B10, 12) become simply

$$\ddot{u}_2 = B(q,\alpha)\left\{ x_1 u_2'' + \frac{x_2}{1+\bar{q}^2} u_2^2 \right\}'' \quad , \quad (B.15)$$

where

$$x_1 = -\bar{q}^2(1+q+a\alpha\bar{q}(1+\bar{q}^2)) \quad (B.16)$$
$$x_2 = a\beta(1+\bar{q}^2) \quad .$$

For q=1 we recover the standard Boussinesq equation namely:

$$\ddot{u}_2 = 2u_2'''' + \tfrac{1}{2}(u_2^2)'' \quad (B.17)$$

**Appendix C:** *q-deformed exponential*

The exponential function $\exp(z)$, is also shown to take a q-deformed form. Indeed, from Eq(2.7) we can extract the following prime derivative

$$(z^n)' \equiv (\partial z^n) = (\sum_{i=0}^{n-1} q^i) z^{n-1}, \quad (C.1)$$

and write the exponential $\exp(z)$ as follows

$$\exp(z) \equiv \sum_{n=0}^{\infty} \frac{z^n}{[n]_q!} \quad (C.2)$$

where we define the q-deformed factorial numbers as follows

$$[n]_q! = 1.(q+1).(q^2+q+1)....(q^{n-1}+q^{n-2}+...q+1). \quad (C.3)$$

With this definition, we see clearly, from Eq(C.1-2), that

$$\partial \exp(z) \equiv \exp(z). \quad (C.4)$$

with the observation that

$$\sum_{i=0}^{n-1} q^i = \frac{[n]_q!}{[(n-1)]_q!} \quad (C.5)$$

Note that one can generalize these definitions of the exponential for an arbitrary function $f(z)$ of conformal spin $\tilde{f}$ by exploiting just the results established before.

**Appendix D:** *Variational principle and q-deformed Euler-Lagrange equations*

Consider the q-deformed Liouville action which we can write as

$$S[\varphi] = \int d^2z \left\{ \partial\varphi\bar{\partial}\varphi + \frac{2}{b}\exp(b\varphi) \right\} \quad (D.1)$$

where a and b are q-dependent coefficients which can be determined using dimensional arguments and conservation of the induced conserved current. The variational principle applied to the q-Liouville action S reads as

$$\delta S[\varphi] = 0 \Leftrightarrow \int d^2 z \left\{ \frac{\partial L}{\partial \varphi} \delta\varphi + \frac{\partial L}{\partial(\partial\varphi)} \delta(\partial\varphi) \right\} = 0 \tag{D.2}$$

where the lagrangian is given by $L = \partial\varphi\bar{\partial}\varphi + \frac{2}{b}\exp(b\varphi)$ with $(\partial = \partial_q)$ and where the variation $\delta$ is required to satisfy $[\delta, \partial]_q = 0$ which means that $\partial\delta = \delta\partial$. Using these remarks and the fact that (by virtue of Eq(2.3),

$$\partial\left(\frac{\partial L}{\partial(\partial\varphi)}\delta\varphi\right) \equiv \left(\frac{\partial L}{\partial(\partial\varphi)}\delta\varphi\right)' = \left(\frac{\partial L}{\partial(\partial\varphi)}\right)'\delta\varphi + \bar{q}^x \frac{\partial L}{\partial(\partial\varphi)}\partial(\delta\varphi) \tag{D.3}$$

whre $x$ is the conformal dimension of $\left(\frac{\partial L}{\partial(\partial\varphi)}\right)$, we obtain the following q-deformed Euler-Lagrange equation

$$\frac{\partial L}{\partial \varphi} - q^x \partial \frac{\partial L}{\partial(\partial\varphi)} = 0 \tag{D.4}$$

for the q-Liouville lagrangian density $L = \partial\varphi\bar{\partial}\varphi + \frac{2}{b}\exp(b\varphi)$. Performing simply algebraic computations, we find

$$\frac{\partial L}{\partial \varphi} = \frac{2}{b}\frac{\partial e^{b\varphi}}{\partial \varphi} \underset{see\ appendix\ C}{=} 2e^{b\varphi}$$
$$\partial \frac{\partial L}{\partial(\partial\varphi)} = \partial\bar{\partial}\varphi \tag{D.5}$$

from which we easily derive the following q-deformed Liouville equation of motion

$$2e^{b\varphi} - q^x \partial\bar{\partial}\varphi = 0 \tag{D.6}$$

On the other hand, using dimensional arguments, we remark that $x = 1$ as the conformal dimension of the lagrangian is $\tilde{L} = 2$. To determine the coefficient constant b we use the conservation of the q-deformed current Eq(5.8) namely

$$T(\varphi) = q\partial^2\varphi - q(\partial\varphi)^2 \quad . \tag{D.7}$$

We have

$$0 = \bar{\partial}T(\varphi) = q\partial(\partial\bar{\partial}\varphi) - q\bar{\partial}(\partial\varphi)^2 \tag{D.8}$$

which fixes the value of the coefficient $b$ namely $b = (1+\bar{q})$ with $\tilde{\varphi} = 0$ and

$$\bar{\partial}(\partial\varphi)^2 = (1+\bar{q})\partial\varphi\bar{\partial}\varphi \tag{D.9}$$

as shown in previous computations. Finally, we have

$$\partial\bar{\partial}\varphi - 2\bar{q}e^{(1+\bar{q})\varphi} = 0 \quad , \quad (\bar{q} = q^{-1}) \tag{D.10}$$

Setting q=1 one recovers the well known Liouville equation $\partial\bar{\partial}\varphi = 2e^{2\varphi}$ associated to the Liouville lagrangian $L = \partial\varphi\bar{\partial}\varphi + \exp(2\varphi)$.

## Appendix E: *q-deformed commutator and compatibility condition*

The use of the q-deformed commutator *Eq(3.8)* instead of the usual one namely $[L, B] = LB - BL$ which is nothing but the *q = 1* limit of *Eq(3.8)* imply a non trivial consideration of the Lax evolution equation *Eq(3.2)* in term of the two compatibility equations. To be more precise let us recall how these equations give rise to the standard evolution Lax equation ( for *q =1*) for arbitrary Lax pair *L, B*. The compatibility equations are given by the following system of linear equations:

$$L\Psi = \lambda\Psi$$
$$B\Psi = \frac{\partial\Psi}{\partial t} \tag{E.1}$$

We have

$$BL\Psi = B\lambda\Psi = \lambda B\Psi = \lambda\frac{\partial\Psi}{\partial t} = \frac{\partial\lambda\Psi}{\partial t} = \frac{\partial L\Psi}{\partial t} \tag{E.2}$$

which gives also

$$BL\Psi = \frac{\partial L\Psi}{\partial t} = \frac{\partial L}{\partial t}\Psi + L\frac{\partial\Psi}{\partial t} = \frac{\partial L}{\partial t}\Psi + LB\Psi \tag{E.3}$$

we have then

$$[B, L]\Psi = (BL - LB)\Psi = \frac{\partial L}{\partial t}\Psi \Leftrightarrow [B, L] = \frac{\partial L}{\partial t} \tag{E.4}$$

In the q-deformed case, the situation is not trivial, since the commutator indispensable to ensure this compatibility is q-deformed. In fact let us consider for simplicity the q-differential Lax pairs $L_2$ and $H_{2n+1} = (L_2^{\frac{2n+1}{2}})_+$ required to satisfy by analogy the Lax evolution equation Eq(3.2)

$$\frac{\partial L_2}{\partial t_{2n+1}} = [H_{2n+1}, L_2]_q \tag{E.5}$$

where the q-deformed commutator is defined in Eq(3.8). As suspected, performing simply algebraic computations, we obtain

$$[H_1, L_2]_q = H_1 L_2 - \bar{q}^2 L_2 H_1 + (\bar{q}^2 - 1)\partial^3 + ....  \quad (E.6a)$$

$$[H_3, L_2]_q = H_3 L_2 - \bar{q}^6 L_2 H_3 + (\bar{q}^6 - 1)\partial^5 + ....  \quad (E.6b)$$

$$[H_5, L_2]_q = H_5 L_2 - \bar{q}^{10} L_2 H_5 + (\bar{q}^{10} - 1)\partial^7 + ....  \quad (E.6c)$$

$$[H_7, L_2]_q = H_7 L_2 - \bar{q}^{14} L_2 H_7 + (\bar{q}^{14} - 1)\partial^9 + ....  \quad (E.6d)$$

.......

results which can be generalized for arbitrary order $n$ of the q-KdV hierarchy as follows

$$[H_{2n+1}, L_2]_q = H_{2n+1} L_2 - \bar{q}^{2(2n+1)} L_2 H_{2n+1} + (\bar{q}^{2(2n+1)} - 1)\partial^{2n+3} + ....  \quad (E.6e)$$

The terms $((\bar{q}^{2(2n+1)} - 1)\partial^{2n+3} + ....)$ in *(E.6)* are extra non linear q- differential operators proportional to $(\bar{q} - 1)$. These extra terms vanish in the standard limit $\bar{q} = 1$ to give rise to the standard commutator *(E.4)*

$$[H_{2n+1}, L_2]_q = H_{2n+1} L_2 - L_2 H_{2n+1}  \quad (E.7)$$

The important remark at this level is that if the compatibility equations exist they must be highly non linear with a dependence in $\bar{q}$ as they should take into account the presence of the non linear extra terms in the q-deformed commutators *(E.6)*. The possibility to write the two compatibility linear equations can emerge naturally as a $\bar{q} = 1$ limit of the previous equations.

### *Acknowledgements*
The authors would like to thank the PARS program N 372/98 CNR who supported this research work. One of the authors M.B.S. would like to thank the good hospitality of the Abdus-Salm ICTP where a big part of this work has been done. He acknowledges also the help of the office of associate and federation schemes and the scientific help of the high-energy section. M.B.S. would like also to acknowledge the constant help of the Faculty of Science at Kenitra and the Department of Physics.

## References:

[1] B. Kupershmidt, Integrable and super-integrable systems (World scientific, Singapore 1990);

   A. Das, Integrable Models (World scientific, Singapore, 1989)

   L.D. Faddeev and L. Takhtajan, Hamiltonian methods in the theory of solitons (Springer Berlin,87)

   P.D.Lax, Commun. Pure Appl. Math. 21 (1968)476 and 28 (1975)141

   M. Jimbo and T. Miwa, Integrable systems in statistical mechanics, World Scientific 1990

[2] Yu.I.Manin and A.O.Radul, Cummun. Math. Phys. 98(1985)65

   K.Yamagishi, Phys.Lett.B.205 (1988) 466

P.Mathieu, Phys.Lett.B.208,101 (1988); Jour.Math. Phys 29 (1988) 2499;

I.Bakas, Phys Lett.B. 219 (1989) 283; Comm.Math .Phys. 123 (1989) 627

J.D.Smit, Comm. Math. Phys. 128 (1990) 1.

[3] A.B.Zamolodchikov, TMP 65 (1985)1205

V.A.Fateev and A.B.Zamolodchikov, Nucl. Phys. B 280[FS18](1988)6411

P. Bouwknegt and K.Schoutens, Phys. Rep. 233(1992)183 (for a review)

[4] E.H.Saidi, M.B.Sedra and A. Serhani; Phys. Lett. B353(1995)209

E.H.Saidi, M.B.Sedra and A. Serhani; Mod. Phys. Lett. A V10N32(1995)

[5] P. Mansfield, Nucl. Phys.B 208 (1982)277, B222(1983)419

D.Olive and N.Turok, Nucl. Phys. B 257[FS14](1986)277

L. Alvarez-Gaumé and C. Gomez, Topics in Liouville Theory, CERN preprint-Th6175/91

[6] E.H.Saidi and M.B.Sedra; Jour .Math. Phys. V35N6 (1994) 3190

M.B.Sedra; Jour .Math. Phys. N37 (1996) 3483

[7] J.E. Humphreys, Introduction to Lie algebras and representation theory, Springer Heïdelberg 1972

V. Kac, Advances in Math. 26(1977)8

J.F. Cornwell, Group Theory in Physics, VIII, (1989)

[8] W.Z.Xian, Introduction to Kac-Moody algebras, world Scientific 1991

[9] M. Jimbo, Lett. Math. Phys. 10, **63** (1985); 11, **247**(1986)

L. Faddeev, Les Houches XXXIX, edited by J. Zuber and R. Stora (Elsevier, Amsterdam, 1984)

J. Wess and B. Zumino, CERN-TH-5697/90, 1980.

V.G. Drinfeld, Quantum groups, Proc. Of the International congress of mathematicians,

Berkeley, 1986, American Mathematical Society 798-1987

[10] E.H. Saidi and M.B. Sedra, Class. Quant. Grav. V10 (1993)1937

E.H. Saidi and M.B. Sedra, Int. Jour. Mod. Phys. V9A6(1994) 891

E.H. Saidi and M.B. Sedra, Mod. Phys.Lett. V9N34(1994)3163

M.B. Sedra, Nucl. Phys. B513(1998)709-722

[11] B. Maroufi, M.Nazah and M.B.Sedra , Extended super-KP Hierarchies, string equations

and Solitons.

E.H. Saidi, M.B. Sedra and J. Zerouaoui., Class. Quant. Grav. V12 (1995)1576;

E.H. Saidi, M.B. Sedra and J. Zerouaoui., Class. Quant. Grav. V12 (1995)1705

I. Benkaddour and E.H. Saidi, Class. Quant. Grav. V16 (1999)1793-1804;

[12] P. Di-Francesco, C. Itzykson and J.B.Zuber, Comm. Math. Phys. 140(1991)543.

[13] E. Frankel, q-alg/9511003 and Inter. Math. Res. Notices No. 2 (1996) 55

[14] M.Rachidi, E.H.Saidi and M.B.Sedra Note on the Di-Francesco and al theorem,  ICTP- IC/95/176